\documentclass[preprint,12pt]{elsarticle}

\usepackage{amssymb}
\usepackage{amsmath}

\usepackage{lineno}

\usepackage{hyperref}

\journal{ } 

\begin{document}

\begin{frontmatter}



\title{Graph Neural Networks embedded into Margules model for vapor-liquid equilibria prediction} 

\author[mpi_affiliation]{Edgar Ivan Sanchez Medina\corref{cor1}}
\ead{sanchez@mpi-magdeburg.mpg.de}

\author[mpi_affiliation,ovgu_affiliation]{Kai Sundmacher}

\cortext[cor1]{Corresponding author}

\affiliation[mpi_affiliation]{organization={Process Systems Engineering, Max Planck Institute for Dynamics of Complex Technical Systems},
            addressline={Sandtorstraße 1}, 
            city={Magdeburg},
            postcode={39106}, 
            state={Saxony-Anhalt},
            country={Germany}}

\affiliation[ovgu_affiliation]{organization={Chair for Process Systems Engineering, Otto-von-Guericke University},
            addressline={Universitätsplatz 2}, 
            city={Magdeburg},
            postcode={39106}, 
            state={Saxony-Anhalt},
            country={Germany}}

\begin{abstract}
Predictive thermodynamic models are crucial for the early stages of product and process design. In this paper the performance of Graph Neural Networks (GNNs) embedded into a relatively simple excess Gibbs energy model, the extended Margules model, for predicting vapor-liquid equilibrium is analyzed. By comparing its performance against the established UNIFAC-Dortmund model it has been shown that GNNs embedded in Margules achieves an overall lower accuracy. However, higher accuracy is observed in the case of various types of binary mixtures. Moreover, since group contribution methods, like UNIFAC, are limited due to feasibility of molecular fragmentation or availability of parameters, the GNN in Margules model offers an alternative for VLE estimation. The findings establish a baseline for the predictive accuracy that simple excess Gibbs energy models combined with GNNs trained solely on infinite dilution data can achieve.
\end{abstract}




\end{frontmatter}



\section{Introduction}
\label{sec:introduction}

Modeling vapor-liquid equilibria is essential for the development of most chemical processes. This is because many chemical processes operate under conditions where vapor and liquid phases interact. Although vapor-liquid equilibria is the most frequently measured phase equilibria property \cite{DDBST}, the vast extension of the chemical space makes it impractical to rely solely on experimental measurements. Consequently, predictive methods have been explored for decades to efficiently predict vapor-liquid equilibria \cite{gmehling2019chemical}, thereby aiding in the design of chemical processes and minimizing the need for experimental measurements in early stages of the process design.

 Various predictive models have been developed for estimating vapor-liquid equilibria. Most of these models are based on mechanistic or phenomenological understanding. Examples include the widely used excess Gibbs energy ($G^E$) model UNIFAC \cite{constantinescu2016further} and equations of state like PCP-SAFT \cite{gross2006equation}. Recently, however, the scientific community has begun exploring hybrid approaches that combine mechanistic and data-driven methods to create predictive models, particularly for predicting activity coefficients. Among these models, some have been developed by embedding a machine learning model within a mechanistic framework, while others enforce thermodynamic constraints directly into the model structure and/or during training.

Machine learning models for predicting activity coefficients that have been embedded into mechanistic expressions include the SPT-NRTL model \cite{winter2023spt}, which combines transformers with the NRTL model, UNIFAC 2.0 \cite{hayer2025advancing}, which uses matrix completion methods to compute the binary interaction matrix of the UNIFAC model, and a similar approach within the UNIQUAC model \cite{jirasek2022making}. Additionally, models developed specifically for predicting the infinite dilution regime, such as the Gibbs-Helmholtz Graph Neural Network (GH-GNN) \cite{medina2023gibbs} and matrix (or tensor) completion methods \cite{damay2021predicting, damay2023predicting}, incorporate an expression derived from the Gibbs-Helmholtz equation as the head of the model to introduce the temperature dependency of activity coefficients at infinite dilution.

Examples of machine learning models that incorporate thermodynamic constraints directly into the model structure include the HANNA model \cite{specht2024hanna}, which combines neural networks with explicit thermodynamic constraints to achieve thermodynamically consistent predictions, and the GE-GNN model \cite{rittig2024thermodynamics}, which similarly enforces Gibbs-Duhem consistency within a graph neural network-based model. While these two examples strictly enforce the consistency of the constraints, other ``soft-constraint" approaches have been explored, such as the Gibbs-Duhem informed graph neural network \cite{rittig2023gibbs}.

Both aspects of integrating machine learning with mechanistic modeling are highly relevant in the field of fluid phase equilibria modeling. We anticipate continued advancements in both areas in the coming years, potentially pushing the boundaries of predictive modeling in this domain. In this contribution, we investigate the incorporation of graph neural network models into a mechanistic framework: the extended Margules model. Although the Margules head model is simpler compared to other hybridized machine learning models (e.g., those combined with head models of NRTL \cite{winter2023spt}, UNIQUAC \cite{jirasek2022making}, or UNIFAC\cite{hayer2025advancing}), it offers a foundational perspective for evaluating how relatively simpler approaches perform compared to more complex ones.

This paper is structured as follows. First, the methods are reviewed, providing an overview of the vapor-liquid equilibrium considerations used for the calculations, the extended Margules model, the data sources and cleaning procedure, and a brief introduction to the GH-GNN model employed in this study. Second, the results are presented and discussed, focusing primarily on the prediction of vapor-liquid equilibria in binary systems, with additional analyses for ternary mixtures. Finally, the conclusions and overview are presented.

\section{Methods}
\label{sec:methods}

\subsection{Vapor-liquid equilibria}

Vapor-liquid equilibria (VLE) is reached when the fugacity of each component $i$ in the liquid phase $L$ and the vapor phase $V$ are equal

\begin{equation}
    f_i^L = f_i^V
    \label{eq_equal_fugacities_vle}
\end{equation}

Two main approaches exist for calculating Equation \ref{eq_equal_fugacities_vle}. The first approach utilizes fugacity coefficients, $\varphi_i$, for both phases, typically derived from an equation of state and, in the case of mixtures, a set of mixing rules. The second approach employs the fugacity coefficient for the vapor phase, along with the activity coefficient, $\gamma_i$, and a standard fugacity, $f_i^0$, for the liquid phase. The standard fugacity is calculated from the fugacity coefficient of the pure liquid at saturation conditions, $\varphi^s_i$, the saturation pressure, $P^s_i$, and the Poynting factor, $\text{Poy}_i$. Therefore, by using the second approach, the VLE of mixture components can be calculated as

\begin{equation}
    x_i \gamma_i \varphi_i^s P_i^s \text{Poy}_i = y_i \varphi_i^V P
    \label{eq_vle_general}
\end{equation}

\noindent where, $x_i$ and $y_i$ refer to the liquid and vapor molar fractions, respectively, and $P$ stands for the system's pressure \cite{gmehling2019chemical}. 

At low or moderate pressures, the $\text{Poy}_i$ is typically close to unity, and for non-associating systems, the values of $\varphi_i^s$ and $\varphi_i^V$ are very similar. Therefore, for many practical applications, it is sufficient to approximate Equation \ref{eq_vle_general} with the following expression:

\begin{equation}
    x_i \gamma_i  P_i^s  \approx y_i P
    \label{eq_vle_extended_raoults_law}
\end{equation}

In this work, Equation \ref{eq_vle_extended_raoults_law} was used to estimate the VLE behavior of components in binary and ternary mixtures of small non-ionic compounds. 

We leverage the previously published GH-GNN model \cite{medina2023gibbs} for predicting infinite dilution activity coefficients (IDACs) and use the extended Margules model to extrapolate the predictions from the infinite dilution regime to finite concentrations. An early analysis of this approach was studied by \citet{medina2023solvent} in the context of solvent pre-selection for extractive distillation, but a rigorous analysis of the approach for predicting binary and ternary VLEs was not there presented. 

\subsection{Extended Margules model}

Different mathematical models have been developed to calculate the molar excess Gibbs energy, $g^E$, of a mixture. Since the concept of $g^E$ is constructed from the idea of modeling a correction to the ideal solution model, any valid $g^E$ model should just satisfy the following limiting condition:

\begin{equation}
    g_i^E \rightarrow 0, \quad \text{as} \quad x_i \rightarrow 0
    \label{eq_boundary_condition}
\end{equation}

One of the simplest approaches to this is to assume that $g^E$ is a continuous and sufficiently smooth function, which can be approximated by an n-th order Taylor series expansion. This approximation results in what is known as the extended Margules model \cite{mukhopadhyay1993discussion}.
A detail derivation is provided in the \ref{app1}.

For a binary mixture, the extended Margules model estimates $g^E$ as:

\begin{equation}
    g^E = x_i x_j (x_i w_{ji} + x_j w_{ij})
    \label{eq_ge_binary_final}
\end{equation}

\noindent where, $w_{ji} = \ln{\gamma_{ji}^\infty}$ and $w_{ij} = \ln{\gamma_{ij}^\infty}$. The activity coefficients at finite concentrations can be then calculated as:

\begin{align}
    \ln{\gamma_i} =& 2 w_{ji} x_i x_j + x_j^2 w_{ij} - 2 g^E \\
    \ln{\gamma_j} =& 2 w_{ij} x_j x_i + x_i^2 w_{ji} - 2 g^E
\end{align}

Similar expressions can be obtained for a ternary mixture, which are outline in the \ref{app1}.

\subsection{Data sources}

In this study, we utilized the open-source GH-GNN model \cite{medina2023gibbs}, trained on experimental IDAC data compiled in Volume IX of the DECHEMA Data Chemistry Series \cite{dechema}. Additionally, experimental binary VLE data sourced from the Korean Data Bank \cite{koreandatabank} was incorporated into our analysis. Both isobaric and isothermal VLE measurements were employed. Since Equation \ref{eq_vle_extended_raoults_law} was used to model the VLE behavior, only systems operating at low to medium pressure levels (i.e., $\leq 500$ kPa) were considered.

The CAS-RN and vapor pressure correlation coefficients of the pure compounds were also obtained from the Korean Data Bank \cite{koreandatabank}. The vapor pressure correlation is defined as:

\begin{equation}
    \ln{P_i^s} = A_i \ln{(T)} + \frac{B_i}{T} + C_i + D_i T^2
    \label{eq_vapor_pressure_kdb}
\end{equation}

\noindent where the pressure $P_i^s$ is expressed in kPa and the temperature $T$ in K. Mixtures containing compounds lacking vapor pressure correlation coefficients ($A_i$, $B_i$, $C_i$ and $D_i$), as well as systems with temperatures outside the correlation’s valid range, were excluded from the analysis.

Additionally, 10 subsets of VLE data for ternary mixtures were collected from the literature. Table \ref{tbl:ternary_systems} describes the specifics of each subset. While limited, the inclusion of ternary VLE systems in the present analysis revels the relative performance of the graph neural networks embedded into the Margules model for multi-component systems. These mixtures include a variety of chemical classes, including aromatics, paraffinic hydrocarbons, nitriles, ketones, alcohols, as well as halogenated and sulfur-containing compounds. In total, 539 data points of ternary mixtures were analyzed.

\begin{table}[h]
\centering
  \caption{Ternary vapor-liquid equilibria data from literature analyzed in this work.}
  \label{tbl:ternary_systems}
  \small
  \begin{tabular*}{1\textwidth}{@{\extracolsep{\fill}}lccccl}
    \hline
    \textbf{Mixture (Component 1/2/3)} & \textbf{\# points} & \textbf{Condition} & \textbf{Ref.} \\
    \hline
    Acetone/Chloroform/Methanol          & 71  & 101.325 kPa  & \cite{hiak1994vapor} \\
    Hexane/Benzene/Sulpholane            & 14  & 101.325 kPa  & \cite{rawat1980isobaric} \\
    Benzene/Heptane/Dimethylformamide    & 50  & 101.325 kPa  & \cite{blanco2000vapor} \\
    Benzene/Heptane/Acetonitrile         & 12  & 101.325 kPa  & \cite{tripathi1975isobaric} \\
    Acetone/Chloroform/Benzene           & 53  & 101.325 kPa  & \cite{kojima1991isobaric} \\
    Benzene/Cyclohexane/Hexane           & 108 & 101.325 kPa  & \cite{ridgway1967physical} \\
    Acetone/Tetrachloromethane/Benzene   & 57  & 101.325 kPa  & \cite{subbarao1966isobaric} \\
    Ethanol/Benzene/Heptane              & 50  & 53.329 kPa   & \cite{nielsen1959vapor} \\
    Hexane/Methanol/Acetone              & 54  & 313.15 K     & \cite{oracz1995vapour} \\
    Chloroform/Methanol/Benzene          & 70  & 101.325 kPa  & \cite{kurihara1998vapor} \\
    \hline
  \end{tabular*}
\end{table}

\subsection{Data cleaning}

All pure components were encoded using SMILES strings retrieved either from the Korean Data Bank \cite{koreandatabank} directly or from PubChem \cite{pubchem} based on their corresponding CAS-RN. When isomeric information was available, isomeric SMILES were used; otherwise, canonical SMILES were employed. If a CAS-RN was not available from the Korean Data Bank \cite{koreandatabank}, SMILES strings were obtained using the OPSiN web tool \cite{opsin}. All data points with compounds for which the SMILES could not be determined were excluded. In total, the SMILES for 110 compounds could not be determined.

The VLE data underwent pre-processing to eliminate entries with decimal point misplacements or inaccurate composition values, such as those where the total molar fractions exceeded unity. In some cases the indexing of compounds with respect to the VLE data was reversed, this was also corrected. VLEs with suspected errors were excluded. All datasets were standardized to consistent units of measurement. Only measurements that provided both liquid and vapor molar fractions were included. Additionally, systems containing molecules with structural features not compatible with the molecular graph construction criteria used to train the GH-GNN \cite{medina2023gibbs} were excluded from the analysis.

The chemical classification of pure compounds from the Korean Data Bank was utilized in this work to evaluate the performance of the proposed framework across different mixture classes. Additionally, following the usage recommendations of the GH-GNN model \cite{medina2023gibbs}, only systems containing compounds with a Tanimoto similarity metric greater than 0.4 were considered. The specifics of this metric are detailed in the original publication \cite{medina2023gibbs} and are intended to control the expected prediction error of the GH-GNN model when applied to systems outside of its training domain.

The resulting VLE dataset of binary mixtures comprises 27,610 data points, involving 235 distinct compounds across 945 unique binary combinations. In total, 781 subsets are isobaric, 903 are isothermal, and an additional 51 subsets were collected under varying conditions. The dataset spans pressures from 0.01 to 499.50 kPa and temperatures ranging from 127.59 to 576.93 K.
 
\subsubsection{Gibbs-Helmholtz Graph Neural Network (GH-GNN)}

The GH-GNN model \cite{medina2023gibbs} represents molecules as graphs following the framework of \citet{battaglia2018relational}, incorporating node-, edge-, and global-level features. These vectorized features encode specific molecular information: atoms are described at the node level, chemical bonds at the edge level, and properties such as polarity and polarizability at the global level. This model is tailored to predict IDACs of small non-ionic compounds. However, extensions of this model have been investigated to handle polymer solutions \cite{sanchez2023gibbs}, ionic liquids \cite{medina2024systematic} and deep eutectic solvents \cite{morales2024graph}.

Graphs representing solutes and solvents are processed using a molecular-level graph neural network (GNN). This GNN updates the graph information in three sequential steps. 

First, the edge features between each pair of nodes $v$ and $w$ are updated according to:

\begin{equation}
\label{eqn:edge_update}
\mathbf{b}_{v,w}^{(l+1)} = \phi_b^{(l)}
\left(
\mathbf{a}_{v}^{(l)} \parallel \mathbf{a}_{w}^{(l)} \parallel \mathbf{b}_{v,w}^{(l)} \parallel \mathbf{u}^{(l)}
\right)
\end{equation}

\noindent where $\mathbf{b}_{v,w}^{(l+1)}$ is the updated edge feature vector, $\mathbf{a}_v^{(l)}$ and $\mathbf{a}_w^{(l)}$ are the feature vectors of nodes $v$ and $w$, respectively, $\mathbf{u}^{(l)}$ represents the global feature vector. The symbol $\parallel$ represents concatenation.

Second, after updating all edges in the graph, the node features are updated according to Equation \ref{eqn:node_update}.

\begin{equation}
   \widehat{\mathbf{b}}_{v}^{(l)} = \sum_{w \in \mathcal{N}(v)} \mathbf{b}_{v,w}^{(l+1)}
\end{equation}

\begin{equation}
\label{eqn:node_update}
   \mathbf{a}_v^{(l+1)} = \phi_a^{(l)}  
   \left( 
        \mathbf{a}_{v}^{(l)} \parallel \widehat{\mathbf{b}}_{v}^{(l)} \parallel \mathbf{u}^{(l)}
   \right)
\end{equation}

\noindent where $\widehat{\mathbf{b}}_{v}^{(l)}$ stands for the aggregated edge information of all edges connecting node $v$ to its neighbors $w \in \mathcal{N}(v)$.

Third, the global feature vector $\mathbf{u}$ is updated according to:

\begin{equation}
   \widetilde{\mathbf{a}}^{(l)} = \frac{1}{n_a} \sum_{v \in \mathcal{V}} \mathbf{a}_v^{(l+1)}
\end{equation}

\begin{equation}
   \widetilde{\mathbf{b}}^{(l)} = \frac{1}{n_b} \sum_{e \in \mathcal{E}} \mathbf{b}_{e}^{(l+1)}
\end{equation}

\begin{equation}
\label{eqn:global_update}
   \mathbf{u}^{(l+1)} = \phi_u^{(l)}  
   \left( 
        \mathbf{u}^{(l)} \parallel \widetilde{\mathbf{a}}^{(l)} \parallel \widetilde{\mathbf{b}}^{(l)}
   \right)
\end{equation}

\noindent where $\widetilde{\mathbf{a}}^{(l)}$ and $\widetilde{\mathbf{b}}^{(l)}$ represent the average node and edge embeddings, respectively. The sets $\mathcal{V}$ and $\mathcal{E}$ stand for the sets of nodes and edges in the graph, and $n_a$ and $n_b$ represent the cardinality of such sets, respectively.

The update functions for the edges $\phi_b^{(l)}$, nodes $\phi_a^{(l)}$ and global features $\phi_u^{(l)}$ are all implemented as single hidden-layer neural networks with the ReLU activation function.

After performing 2 sequential message passing transformations (described by Equations \ref{eqn:edge_update} to \ref{eqn:global_update}), the final graphs are pooled into vector embeddings. These embeddings are then combined to form a new graph representing the mixture, where nodes correspond to chemical species and edges denote hydrogen-bonding interactions. This mixture graph is subsequently passed through a second, mixture-level GNN, and the output is pooled into a vector representing the mixture. This vector embedding is finally used to estimate temperature-independent parameters in an expression derived from the Gibbs-Helmholtz equation, enabling the prediction of IDACs.

In this work, the GH-GNN model is embedded into the extended Margules model. This combination allows the GH-GNN to predict the parameters of the extended Margules model, corresponding to the respective binary IDACs, which are then used to calculate finite concentration activity coefficients. By embedding the GH-GNN into the Margules framework, we ensure that the predictions remain Gibbs-Duhem consistent. In the following sections, we analyze the performance of this combined model in predicting vapor-liquid equilibria and compare its results with those of the widely used UNIFAC-Dortmund \cite{constantinescu2016further} model.

\section{Results and discussion}
\label{sec:results_and_discussion}

\subsection{Isothermal binary vapor-liquid equilibria}

The 903 isothermal subsets from the cleaned dataset were predicted using the GH-GNN model embedded within the extended Margules model. In total, 12,980 data points were analyzed, involving 180 distinct compounds across 531 unique binary combinations. The temperature range of the isothermal subsets spans from 127.59 K to 573.15 K.

\begin{figure}[h]
    \centering
    \includegraphics[width=1\textwidth]{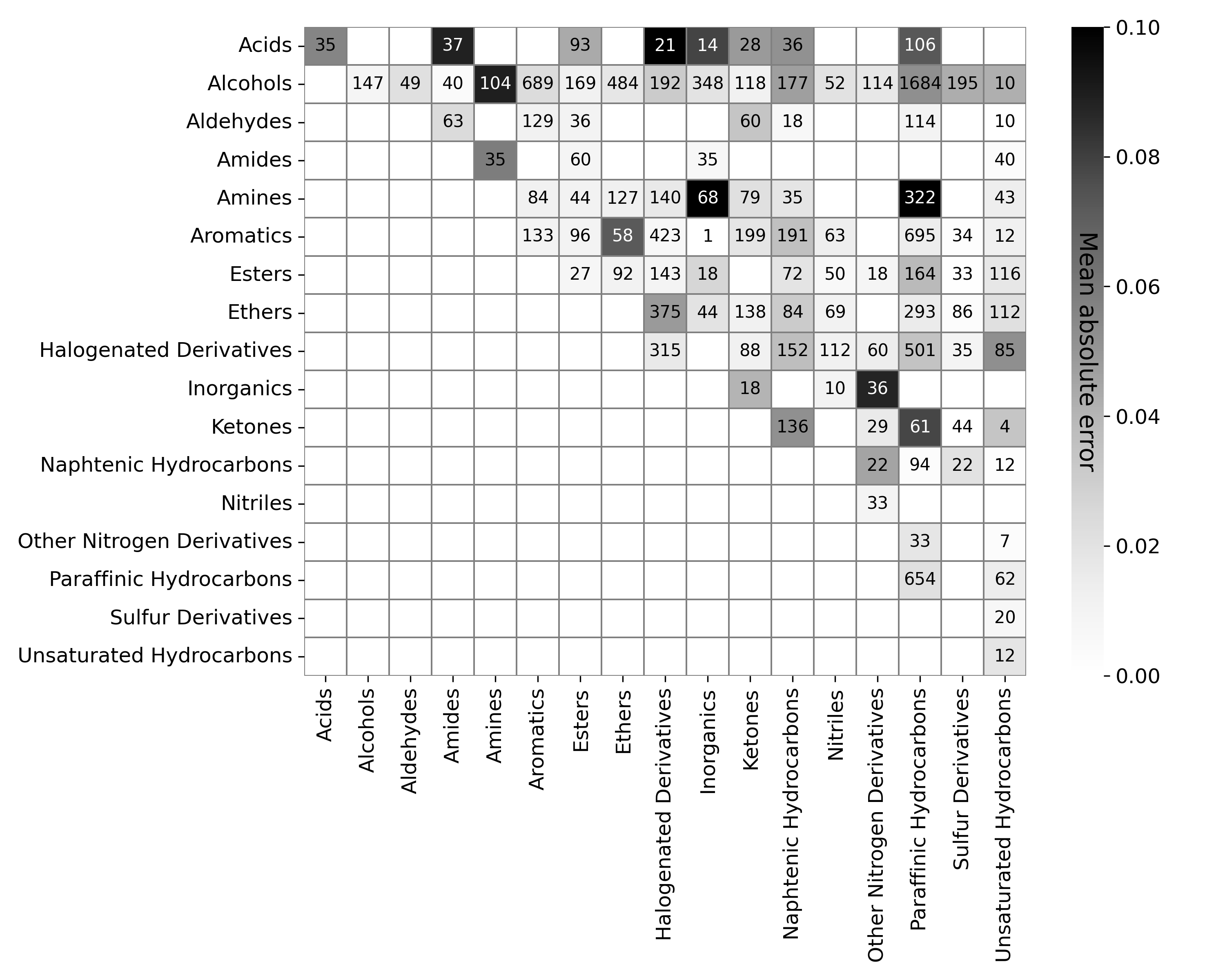}
    \caption{Heatmap of the mean absolute error achieved by the GH-GNN model embedded into the Margules model for different types of binary mixtures at isothermal conditions. The error is measured with respect to the vapor phase molar fraction. The number in each cell represents the number of data points included in the corresponding binary category.}
    \label{fig:isothermal_map}
\end{figure}

To compute the activity coefficients for all isothermal systems in the dataset, a total of 1062 IDACs were required. Of the 531 binary systems, 120 (22.6\%) had both IDACs observed during training. For 382 systems (72\%), the GH-GNN model had to predict at least one of the two required IDACs in other solute-solvent combination as the ones observed during training. For the remaining 29 systems (5.4\%), the model needed to predict at least one of the necessary IDACs with molecules never observed before during training. This was due to 15 compounds (out of the 180 in the isothermal VLE data) not being present in the GH-GNN training set. The fact that even for this limited VLE dataset only very few of the necessary IDACs have been measured experimentally underscores the strong motivation for developing reliable predictive methods.

\begin{figure}[h]
    \centering
    \includegraphics[width=0.85\textwidth]{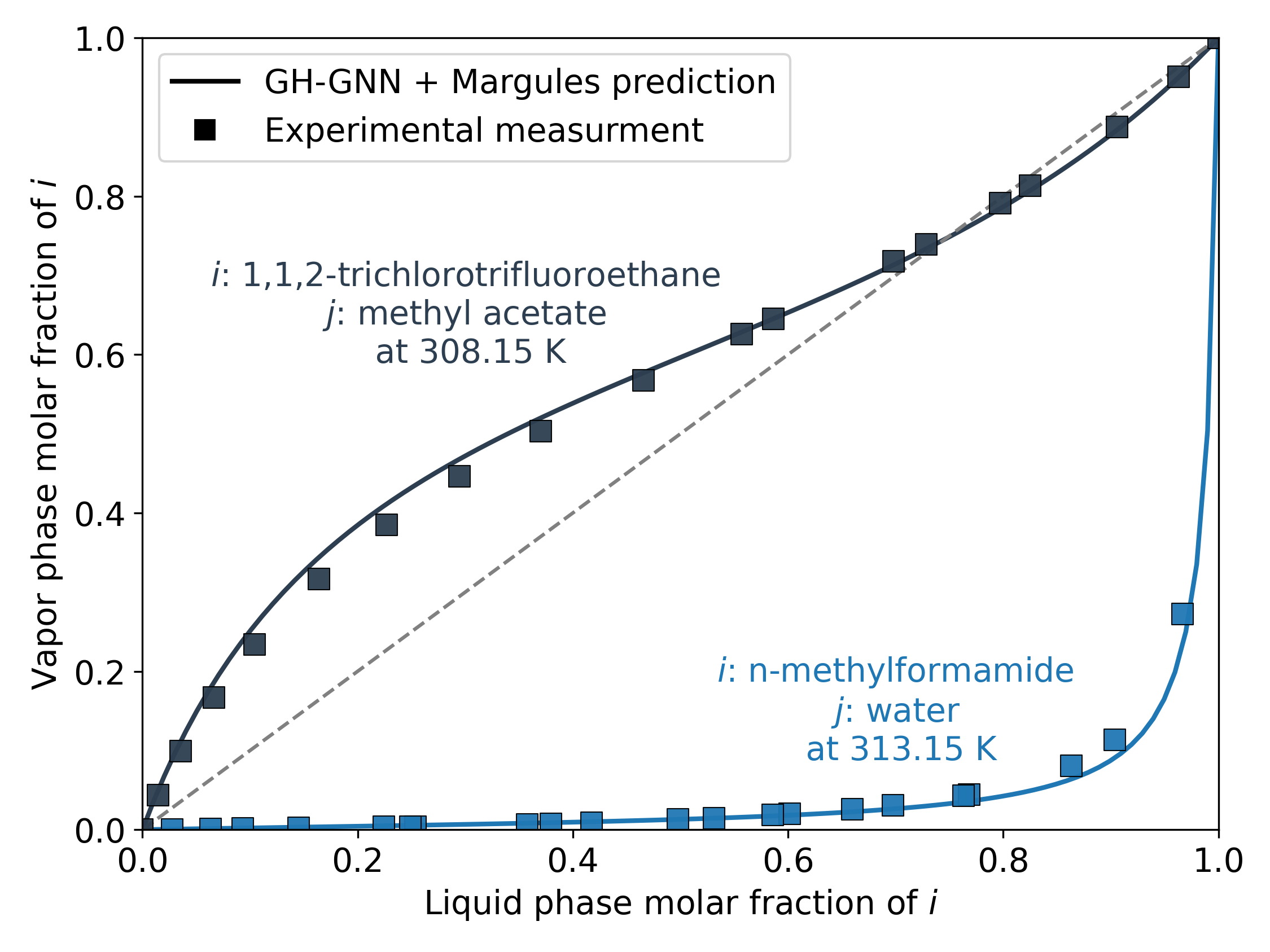}
    \caption{Isothermal vapor-liquid equilibria (VLE) diagram for two systems that are unfeasible to predict with UNIFAC-Dortmund.}
    \label{fig:isothermal_diagram}
\end{figure}

A similar analysis for UNIFAC is challenging to perform, as the specific mixtures and data used for its parametrization are not readily accessible. Furthermore, it is important to note that while the GH-GNN embedded within the Margules model had access exclusively to IDAC data, UNIFAC was parameterized using a broader range of experimental data, including IDACs, vapor-liquid equilibria (VLE), liquid-liquid equilibria (LLE), solid-liquid equilibria (SLE), azeotropic data, and calorimetric measurements \cite{gmehling2002modified}.

Figure \ref{fig:isothermal_map} presents a heatmap of the mean absolute error (MAE) achieved by the GH-GNN embedded in the Margules model when predicting the vapor-phase molar fraction of the binary vapor-liquid equilibria (VLE) under isothermal conditions. 

Overall, systems containing acids are particularly challenging for the analyzed model to predict, especially when combined with halogenated species or amides. This difficulty likely arises from the strong hydrogen-bonding and dipole-dipole interactions characteristic of these compounds, which may be inadequately captured by the information provided to the GH-GNN model. Additionally, the Margules expression might require higher-order polynomial terms to better account for these complex interactions. Similarly, systems involving alcohols and amines also exhibit poor predictive accuracy. Furthermore, systems containing water (categorized here under the ``Inorganic" class), particularly in combination with amines and other nitrogen derivatives, were associated with relatively large prediction errors.

\begin{table}[h]
\centering
  \caption{Comparison between UNIFAC-Dortmund and the GH-GNN model embedded into the Margules model for predicting binary vapor-liquid equilibria (VLE) under isothermal conditions. The comparison is according to the mean absolute error (MAE), the coefficient of determination (R$^2$) and the percentage of points predicted with an absolute error of less than or equal to 0.03. All metrics are with respect to the predicted and actual values of the molar fraction in the vapor phase.}
  \label{tbl:isothermal}
  \small
  \begin{tabular*}{1\textwidth}{@{\extracolsep{\fill}}lccc}
    \hline
    \textbf{Model} & \textbf{MAE} $\downarrow$ & \textbf{R$^2$} $\uparrow$ & \textbf{AE $\leq 0.03$} $\uparrow$ \\
    \hline
    UNIFAC-Dortmund & 0.0214 & 0.957 & 86.72\% \\
    GH-GNN + Margules & 0.0307 & 0.948 & 74.77\% \\
    \hline
  \end{tabular*}
\end{table}

The comparison between the UNIFAC-Dortmund and GH-GNN model embedded into the Margules model, presented in Table \ref{tbl:isothermal}, indicates that UNIFAC-Dortmund predicts most isothermal VLE binary systems more accurately than the GNN-based Margules hybrid model. This comparison includes all mixtures that can be feasibly predicted using the UNIFAC model (12,700 data points, 514 binary mixtures).

However, this result is not unexpected, given that, as previously mentioned, extensive VLE data were also used in the parametrization of UNIFAC-Dortmund. In contrast, the GH-GNN model, combined with the Margules framework, achieves its performance using only experimental IDAC data. Consequently, while the GNN-based predictive model performs some degree of extrapolation from the infinite regime to finite concentrations, the UNIFAC model is likely reproducing VLE data that were part of its original parametrization. Overall, this comparison shows the predictive limitations of GNNs embedded into the Margules framework, but it also establishes a baseline of the degree of accuracy that can be reached by employing scarce data at infinite dilution with GNN models and Gibbs-Duhem consistent expressions for predicting VLEs.

When comparing the flexibility of the two predictive approaches, UNIFAC-Dortmund and the GNNs embedded in the Margules model, some systems in the dataset cannot be predicted using the UNIFAC model. This limitation arises either from the unfeasibility of fragmentation or the lack of available interaction parameters. In this regard, the proposed framework combining GNNs with the Margules model offers an effective alternative.

For example, Figure \ref{fig:isothermal_diagram} shows the VLE diagram for two systems that UNIFAC-Dortmund cannot predict. In both cases, at least one of the necessary IDACs was not available experimentally, and the GH-GNN model was needed for prediction. Notably, as shown for the system ``1,1,2-trichlorotrifluoroethane / methyl acetate" in Figure \ref{fig:isothermal_diagram}, the GH-GNN model embedded within Margules successfully captures azeotropic behavior with significant precision, even without dedicated azeotropic data used during training.

\subsection{Isobaric binary vapor-liquid equilibria}

The 781 isobaric subsets contain a total of 14,118 data points, covering 162 distinct compounds in 525 different binary combinations. The pressure ranges from 1.33 to 496.63 kPa. Among these subsets, 106 systems (20.2\%) had both necessary IDACs available from experimental measurements and were therefore observed during the GH-GNN model training. Additionally, 409 systems (77.9\%) required interpolation for at least one IDAC, meaning the values were predicted based on observations of the compounds in other binary combinations. The remaining 10 systems (1.9\%) involved molecules that were entirely absent during GH-GNN model training (i.e., extrapolations).

In order to perform the isobaric VLE calculations, an iterative process is needed in which the temperature used for predicting the activity coefficients is changed in order to minimize the difference between the estimated system pressure and the actual pressure. This is represented in the optimization problem \ref{eq_optim_isobaric}, which was here solved using the SciPy package \cite{virtanen2020scipy} with Brent's algorithm, allowing up to 2,000 iterations and a tolerance of $1.48 \times 10^{-8}$.

\begin{equation}
\begin{aligned}
\underset{\substack{T}}{\text{min}} \quad & \vert P - \hat{P}(T) \vert \\
\text{s.t.} \quad & \hat{P}(T) = x_i \gamma_i(T) P_i^{sat}(T) + x_j \gamma_j(T) P_j^{sat}(T) \\
& \{\gamma_i(T), \gamma_j(T)\} \leftarrow \text{GH-GNN embedded in Margules}(T)\\
& T_{\min} \leq T \leq T_{\max}
\end{aligned}
\label{eq_optim_isobaric}
\end{equation}

\noindent where $P$ and $\hat{P}$ denote the actual and predicted system pressures, respectively, and $T_{\min}$ and $T_{\max}$ define the optimization bounds for the temperature. In this case, these bounds correspond to the minimum temperature range in which the vapor pressure correlation (Equation \ref{eq_vapor_pressure_kdb}) remains valid.

\begin{figure}[h!]
    \centering
    \includegraphics[width=1\textwidth]{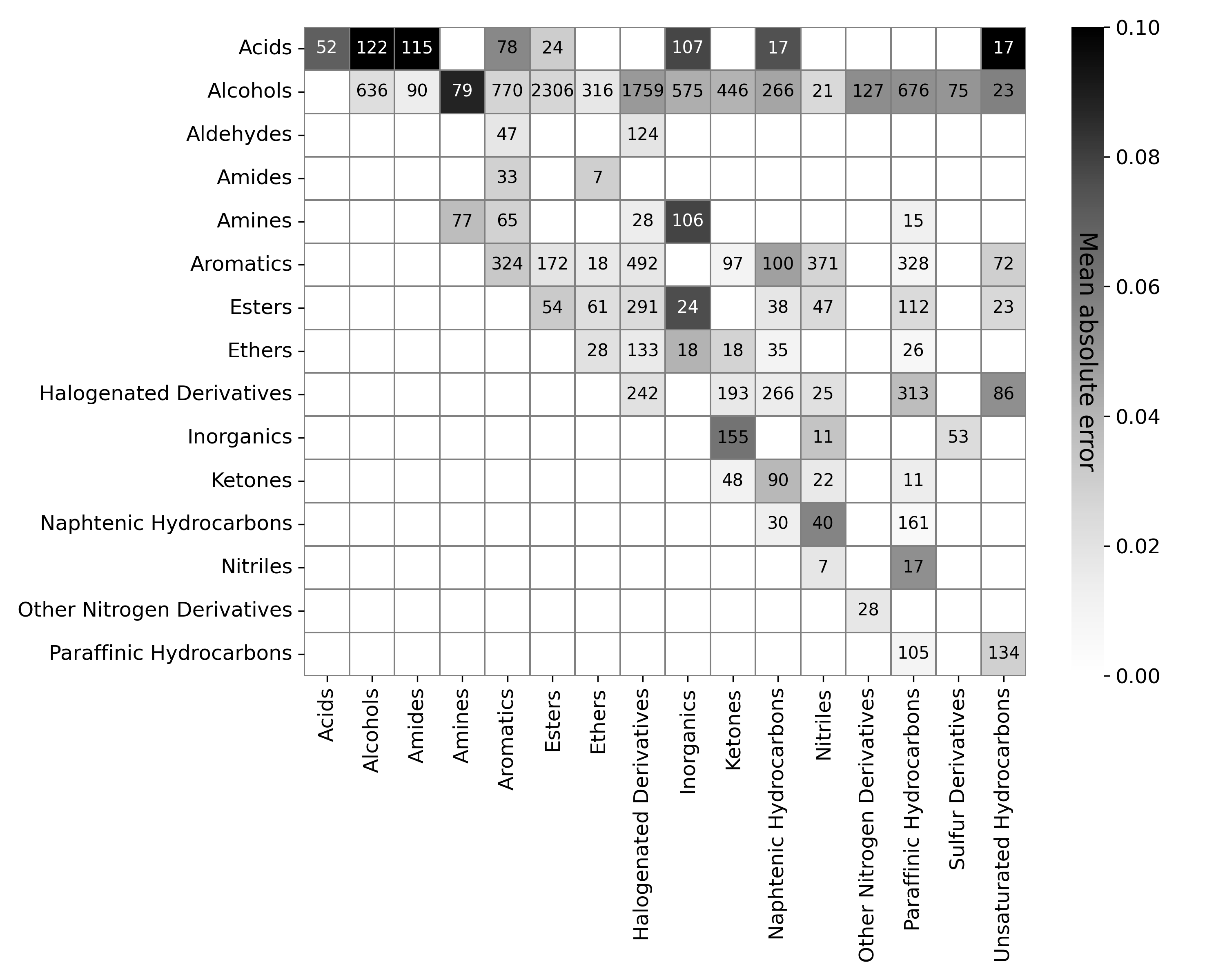}
    \caption{Heatmap of the mean absolute error achieved by the GH-GNN model embedded into the Margules model for different types of binary mixtures at isobaric conditions. The error is measured with respect to the vapor phase molar fraction. The number in each cell represents the number of data points included in the corresponding binary category.}
    \label{fig:isobaric_map}
\end{figure}

\begin{table}[h]
\centering
  \caption{Comparison between UNIFAC-Dortmund and the GH-GNN model embedded into the Margules model for predicting binary vapor-liquid equilibria (VLE) under isobaric conditions. The comparison is according to the mean absolute error (MAE), the coefficient of determination (R$^2$) and the percentage of points predicted with an absolute error of less than or equal to 0.03. All metrics are with respect to the predicted and actual values of the molar fraction in the vapor phase.}
  \label{tbl:isobaric}
  \small
  \begin{tabular*}{1\textwidth}{@{\extracolsep{\fill}}lccc}
    \hline
    \textbf{Model} & \textbf{MAE} $\downarrow$ & \textbf{R$^2$} $\uparrow$ & \textbf{AE $\leq 0.03$} $\uparrow$ \\
    \hline
    UNIFAC-Dortmund & 0.0224 & 0.967 & 81.91\% \\
    GH-GNN + Margules & 0.0350 & 0.946 & 65.78\% \\
    \hline
  \end{tabular*}
\end{table}

Table \ref{tbl:isobaric} compares the performance of the UNIFAC-Dortmund model and the GH-GNN model embedded into Margules. The comparison is shown only for the systems that are feasible to predict with UNIFAC, resulting in 14,075 data points. Similar to the isothermal systems, UNIFAC-Dortmund generally outperforms the GH-GNN model combined with Margules in predicting isobaric VLEs of binary systems. As previously explained, this outcome is expected due to the difference in available experimental data during training. Nevertheless, it provides a useful baseline for assessing the accuracy achievable when predicting finite compositions from infinite dilution information using predictive methods based on GNNs.

\begin{figure}[h!]
    \centering
    \includegraphics[width=0.85\textwidth]{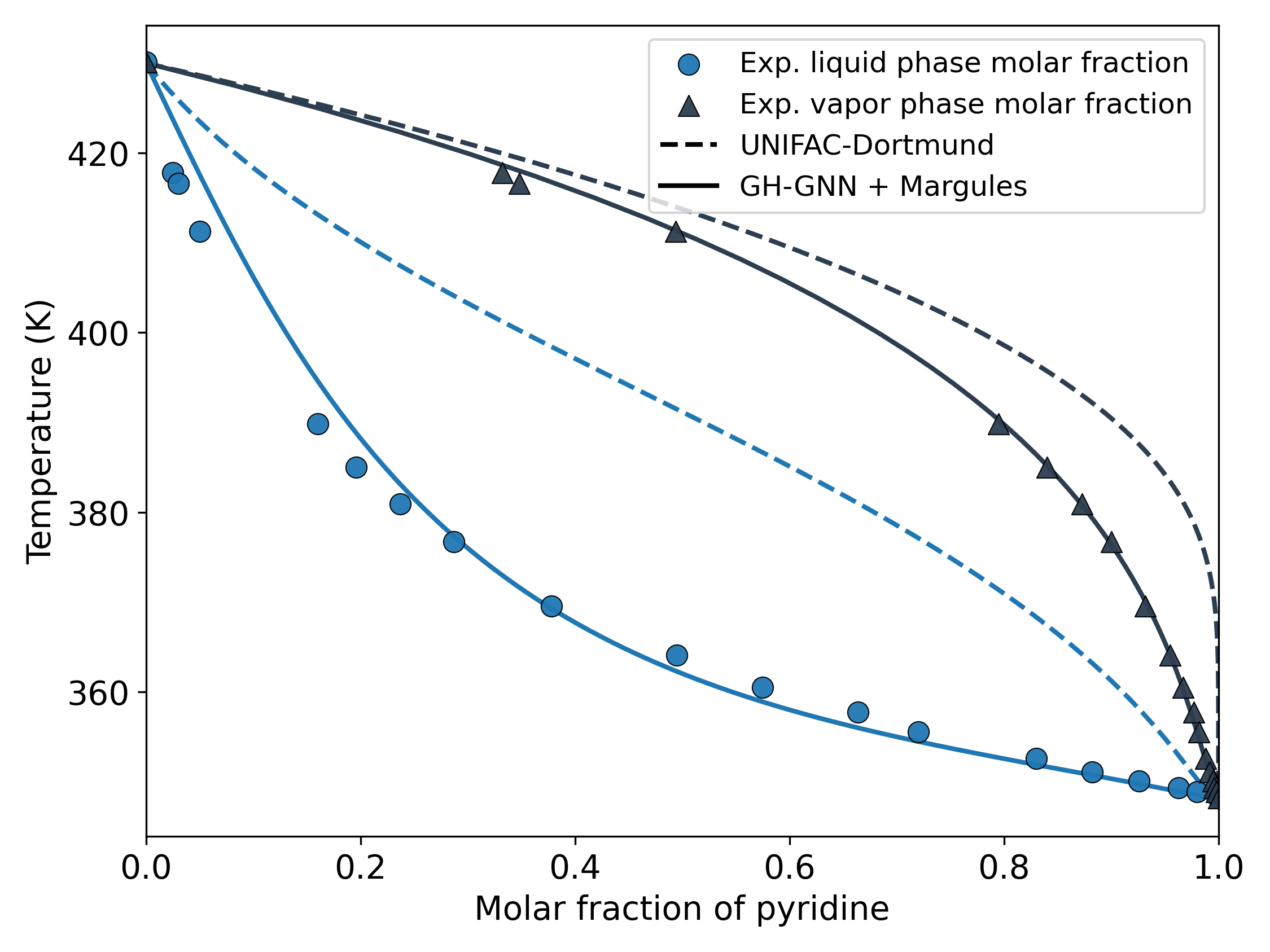}
    \caption{Isobaric vapor-liquid equilibria (VLE) diagram comparing experimental data with the predictions of UNIFAC-Dortmund and the GH-GNN model embedded into the Margules model. The system is pyridine/1,2,3,4-tetrahydronaphthalene at 26.66 kPa.}
    \label{fig:isobaric_diagram_1}
\end{figure}

The heatmap in Figure \ref{fig:isobaric_map} illustrates the performance of the GNN-Margules-based model across different binary combinations. Notably, compared to the isothermal data, different binary classes are now considered. For instance, combinations such as acids/alcohols and acids/aromatics are present under isobaric conditions but absent in the isothermal dataset. In total, 16 new binary classes appear in the isobaric data that were not included in the isothermal data. However, 40 binary classes are present only at isothermal conditions. In total the intersection of binary classes between isothermal and isobaric conditions is 60. Similar to the isothermal case, many binary classes under isobaric conditions are predicted with relatively low errors.

\begin{figure}[h]
    \centering
    \includegraphics[width=0.85\textwidth]{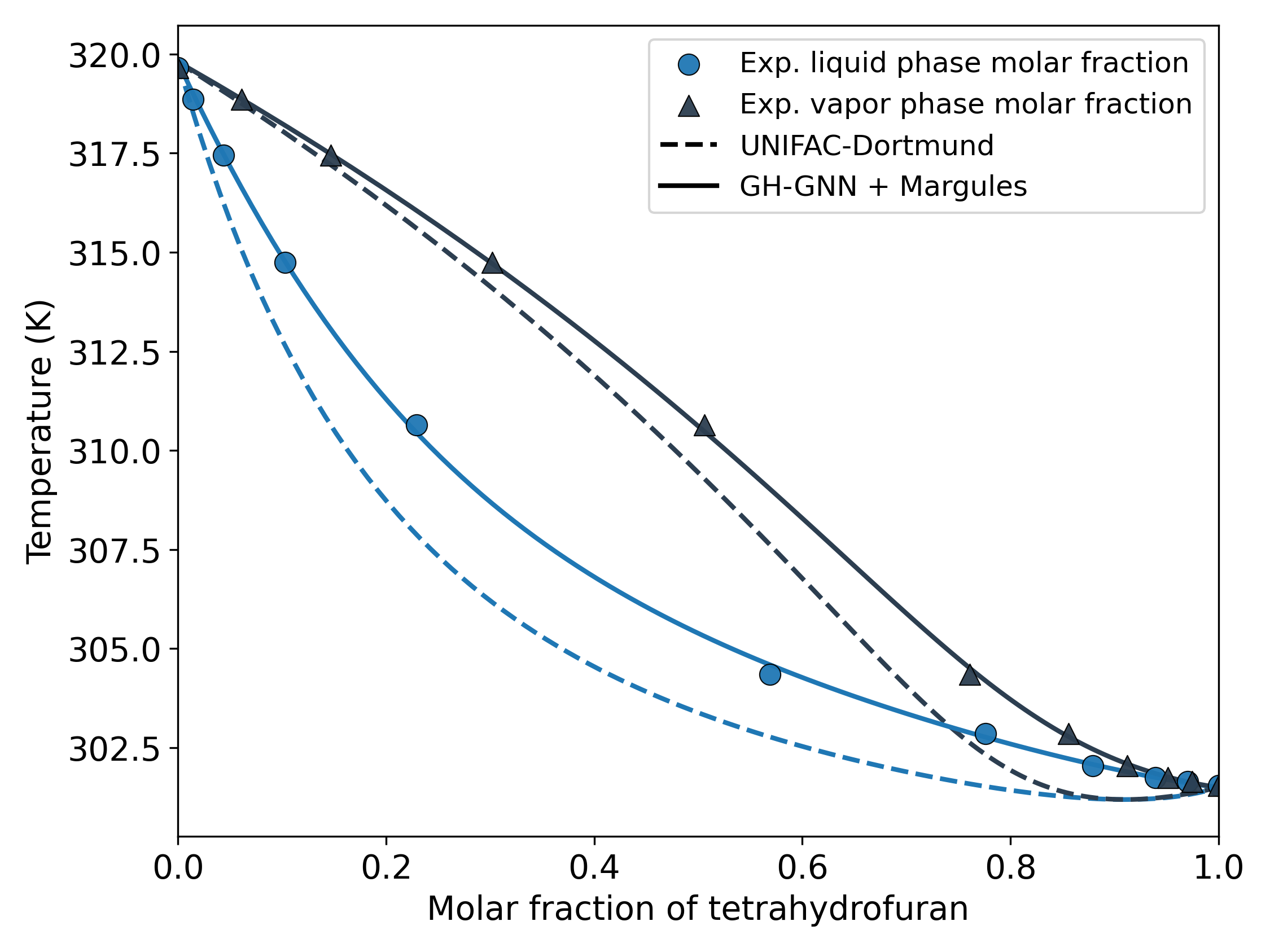}
    \caption{Isobaric vapor-liquid equilibria (VLE) diagram comparing experimental data with the predictions of UNIFAC-Dortmund and the GH-GNN model embedded into the Margules model. The system is tetrahydrofuran/ethanol at 25 kPa.}
    \label{fig:isobaric_diagram_2}
\end{figure}

Out of the 737 isobaric systems that are feasible to predict with UNIFAC-Dortmund, 182 are, on average, better predicted by the GNN-Margules-based model than by UNIFAC-Dortmund. To illustrate this, Figure \ref{fig:isobaric_diagram_1} presents the VLE diagram for the system ``pyridine/1,2,3,4-tetrahydronaphthalene" at 26.66 kPa. It is evident that the proposed model here predicts the VLE behavior more accurately compared to UNIFAC-Dortmund. A similar trend is observed in Figure \ref{fig:isobaric_diagram_2}, which shows the VLE diagram for the system “tetrahydrofuran/ethanol" at 25 kPa.

In these two isobaric VLE examples, the GH-GNN model interpolates at least one of the two required IDACs. This highlights that, although UNIFAC-Dortmund generally outperforms the proposed model, there are counterexamples where the GH-GNN combined with Margules model provides superior predictions, which may be relevant in specific practical scenarios where such mixtures might be of interest.

\subsection{Overall performance predicting binary vapor-liquid equilibria}

\begin{figure}[h!]
    \centering
    \includegraphics[width=1\textwidth]{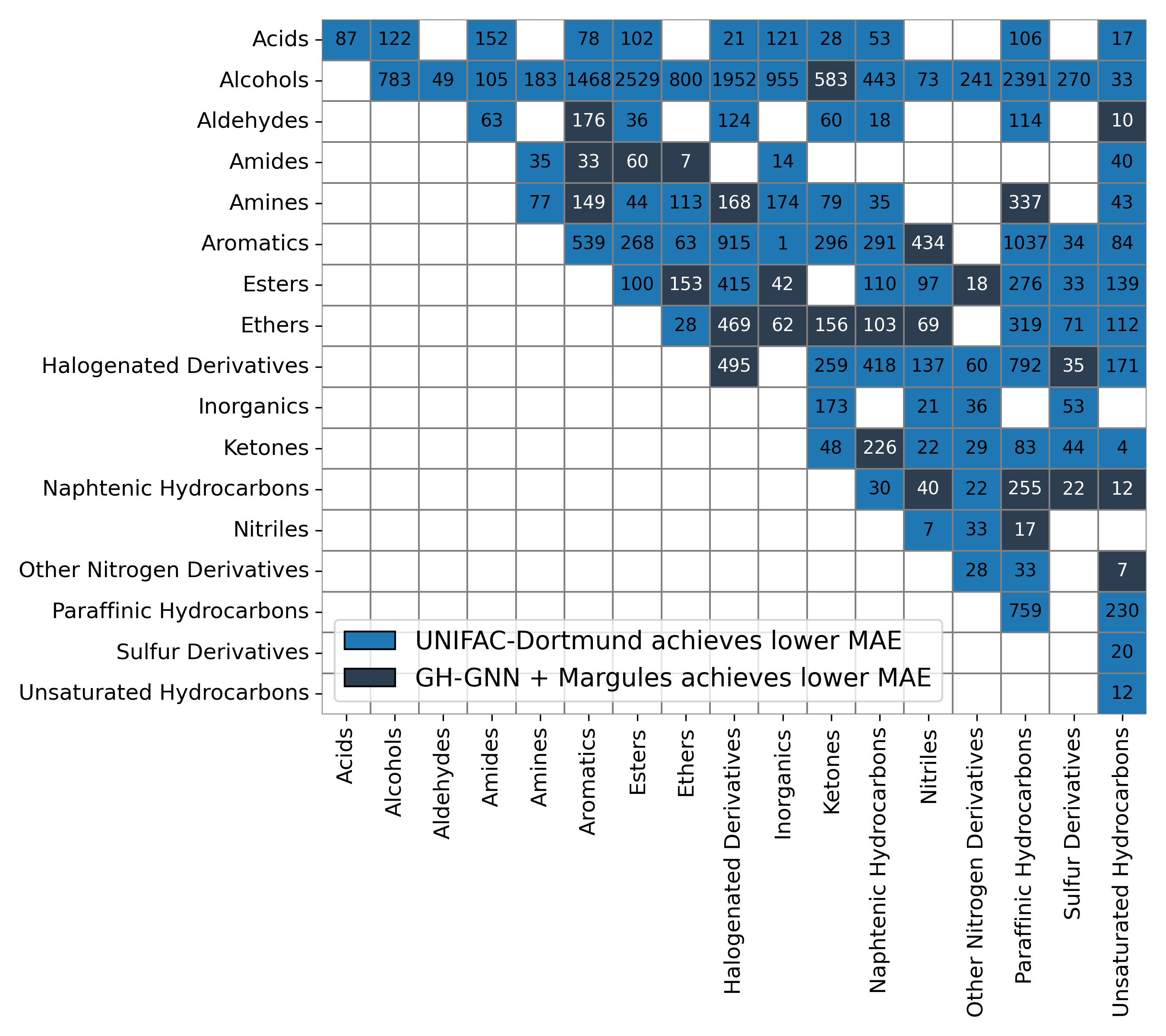}
    \caption{Comparison between the UNIFAC-Dortmund and the GH-GNN in Margules model based on the mean absolute error (MAE) for predicting the vapor molar fraction. All feasible systems at isothermal, isobaric and random conditions are included. The number in each cell represents the number of data points.}
    \label{fig:mae_matrix_ghgnn_vs_unifac}
\end{figure}

To gain a more detailed understanding of the binary classes where the GH-GNN model embedded into Margules outperforms UNIFAC-Dortmund, Figure \ref{fig:mae_matrix_ghgnn_vs_unifac} presents the best-performing model for each binary combination. The performance of each binary class is evaluated based on the mean absolute error (MAE) across all temperature and pressure conditions, including the isothermal, isobaric, and random subsets. Predictions are assessed with respect to the molar fraction in the vapor phase.

\begin{figure}[h]
    \centering
    \includegraphics[width=0.85\textwidth]{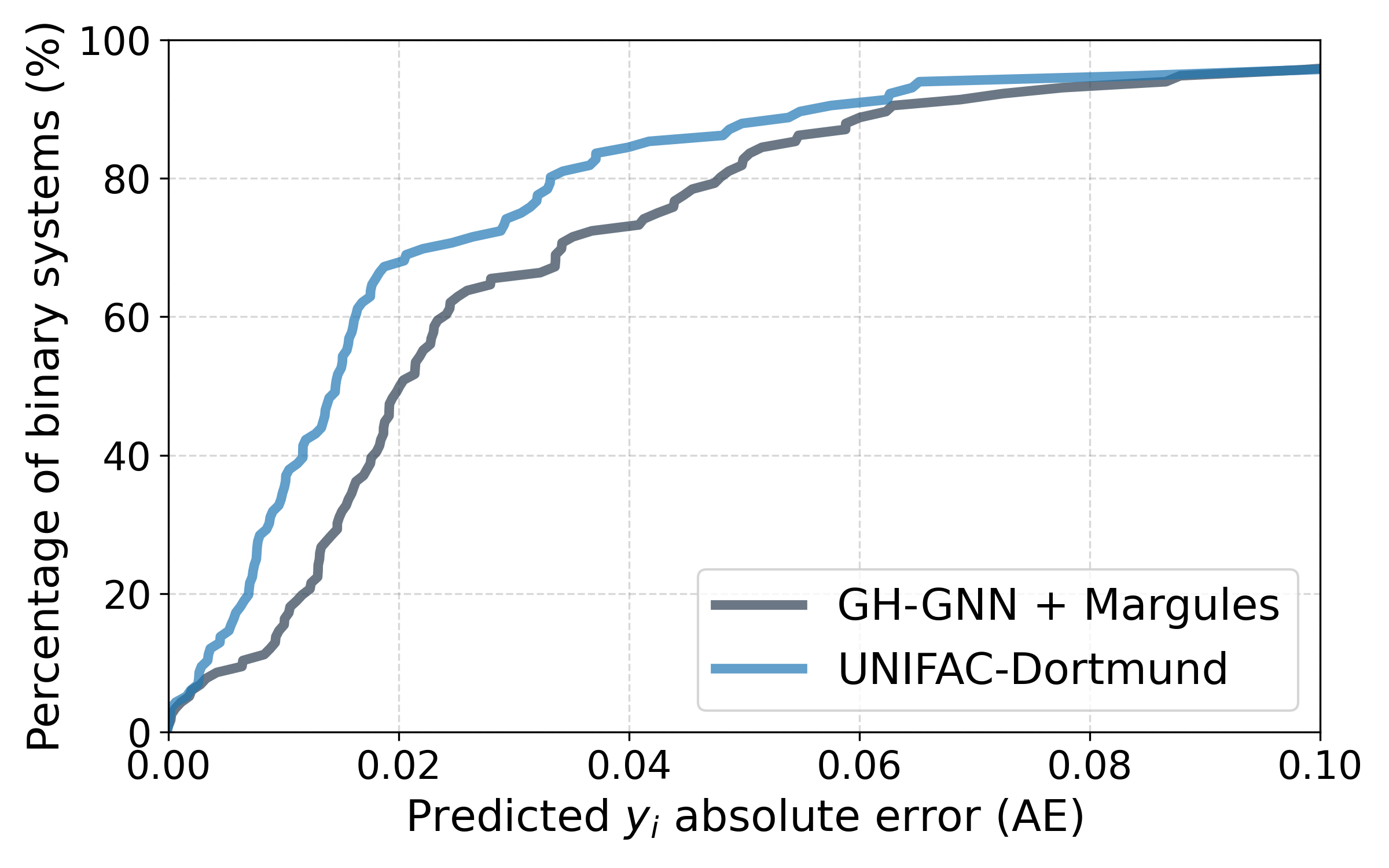}
    \caption{Cumulative percentage of binary vapor-liquid equilibria data points predicted below different thresholds of absolute error with respect to the molar fraction in the vapor phase. The performance is shown for the models UNIFAC-Dortmund and the GH-GNN embedded into Margules.}
    \label{fig:cummulative}
\end{figure}

As previously discussed for the isothermal and isobaric cases, the UNIFAC-Dortmund model generally performs better for most binary classes. However, certain binary classes are predicted more accurately by the GH-GNN model combined with Margules. For instance, the binary class ``amines/aromatics," which includes the system shown in Figure \ref{fig:isobaric_diagram_1}, tends to be better predicted using the proposed GH-GNN in Margules model. Out of the 116 distinct binary classes studied in this work, 27 binary classes are on averaged predicted with lower errors by the GH-GNN in Margules model compared to UNIFAC-Dortmund.

Figure \ref{fig:cummulative} presents the cumulative percentage of binary VLE data points predicted with absolute errors below various thresholds, based on the molar fraction in the vapor phase. Both models achieve a similar percentage of well-predicted data points for absolute errors up to around 0.01. However, UNIFAC-Dortmund consistently yields lower absolute errors overall. This is primarily attributed to differences in the experimental data available during training but also highlights the potential of GNN-based hybrid models for extrapolating finite concentration conditions from limited infinite dilution data. Including VLE data in the training of GNN-based models alongside infinite dilution data would likely lead to a significant increase in accuracy, as already observed in the case of transformers \cite{winter2023spt}.

\subsection{Ternary vapor-liquid equilibria}

In this section, we provide insights into the relative performance of the GH-GNN model embedded in Margules compared to UNIFAC-Dortmund for ternary mixtures. As shown in Table \ref{tbl:ternary_systems}, the available ternary data is much more limited. Therefore, this comparison aims to provide qualitative insights rather than an in-depth analysis, as conducted for binary mixtures.

\begin{table}[h]
\centering
  \caption{Comparison between UNIFAC-Dortmund and the GH-GNN model embedded into the Margules model for predicting ternary vapor-liquid equilibria (VLE). The comparison is according to the mean absolute error (MAE), the coefficient of determination (R$^2$) and the percentage of points predicted with an absolute error of less than or equal to 0.03. All metrics are with respect to the predicted and actual values of the molar fraction in the vapor phase of components 1 and 2.}
  \label{tbl:ternary}
  \small
  \begin{tabular*}{1\textwidth}{@{\extracolsep{\fill}}lccc}
    \hline
    \multicolumn{4}{c}{\textbf{Isothermal system}}\\
    \hline
    \textbf{Model} & \textbf{MAE} $\downarrow$ & \textbf{R$^2$} $\uparrow$ & \textbf{AE $\leq 0.03$} $\uparrow$ \\
    \hline
    UNIFAC-Dortmund & 0.0059 & 0.995 & 100\% \\
    GH-GNN + Margules & 0.0568 & 0.394 & 37.96\% \\
    \hline
    \multicolumn{4}{c}{\textbf{Isobaric systems}}\\
    \hline
    \textbf{Model} & \textbf{MAE} $\downarrow$ & \textbf{R$^2$} $\uparrow$ & \textbf{AE $\leq 0.03$} $\uparrow$ \\
    \hline
    UNIFAC-Dortmund & 0.0123 & 0.991 & 89.07\% \\
    GH-GNN + Margules & 0.0330 & 0.925 & 66.60\% \\
    \hline
  \end{tabular*}
\end{table}

Similar to the case of binary systems presented earlier, UNIFAC-Dortmund generally provides more accurate estimations of ternary VLEs (see Table \ref{tbl:ternary}). Despite this, predictions from the GH-GNN model combined with Margules also align well with the physical behavior of the mixtures. Moreover, as observed in binary mixtures, the GH-GNN embedded in Margules produces more accurate predictions for certain systems compared to UNIFAC-Dortmund.

Specifically, among the 10 ternary systems analyzed, the GH-GNN in Margules achieves a lower MAE than UNIFAC-Dortmund for the following systems: ``benzene/heptane/dimethylformamide" (MAE of 0.0296 vs. 0.0297 for UNIFAC-Dortmund), ``benzene/heptane/acetonitrile" (MAE of 0.0084 vs. 0.0114 for UNIFAC-Dortmund), and ``acetone/chloroform/benzene" (MAE of 0.0047 vs. 0.0184 for UNIFAC-Dortmund). For the remaining systems, UNIFAC-Dortmund achieves lower overall MAE. This exemplifies that, for systems where UNIFAC-Dortmund cannot provide predictions due to missing interaction parameters or unfeasibility of fragmentation into functional groups, the GH-GNN in Margules model offers a useful first-step approximation of the mixture behavior, which could be beneficial in early stages of molecular and process design.

\section{Conclusions}
\label{sec:conclusions}

Predictive models for thermophysical properties of pure compounds and their mixtures play a crucial role in chemical space exploration for product and process design. While numerous predictive methods have been developed over the decades, they remain limited in terms of both accuracy and the range of systems they can effectively describe.

In this work, we studied the performance of a GNN-based model embedded in the extended Margules framework, which assumes that the excess Gibbs energy can be approximated using a Taylor polynomial. The performance of this model was compared against the widely used UNIFAC-Dortmund model across extensive vapor-liquid equilibrium (VLE) data for binary systems under isothermal and isobaric conditions.

Additionally, we provide insights into the relative performance of these models in predicting ternary VLEs. Overall, UNIFAC-Dortmund yields more accurate VLE predictions. However, a crucial factor in this comparison is the difference in the types of experimental data used for model development. While the GNN embedded in Margules was trained exclusively on limited infinite dilution data, requiring extrapolation from infinite dilution to finite concentrations, UNIFAC-Dortmund was parameterized using a much broader dataset, including infinite dilution data, VLE, LLE, SLE, azeotropic, and caloric data \cite{constantinescu2016further,gmehling2002modified}.

It is likely that incorporating VLE data alongside infinite dilution data could significantly enhance the accuracy of GNN-based models within the Margules framework. Nonetheless, this study establishes a valuable baseline, demonstrating the predictive accuracy achievable when extrapolating VLE behavior solely from binary infinite dilution data using a relatively simple excess Gibbs energy model in combination with a highly flexible GNN-based model.

Future work should focus on a comprehensive comparison of GNN-based hybrid models trained on extensive and diverse datasets. However, a key challenge in this endeavor is the structuring and open accessibility of thermodynamic data for model development and benchmarking. Advancing community efforts in thermodynamic data curation and structuring will be essential for the continuous improvement and open-source benchmarking of predictive thermodynamic models.

\section*{Data availability}
The vapor-liquid equilibria (VLE) and pure component data used in this work is available at the accompanying GitHub repository at \url{https://github.com/edgarsmdn/GH_GNN_Margules}.

\section*{Code availability}
The code for generating the figures and running VLE predictions with the GH-GNN model embedded into the Margules expression are available at the accompanying GitHub repository at \url{https://github.com/edgarsmdn/GH_GNN_Margules}.

\section*{CRediT authorship contribution statement}
\textbf{Edgar Ivan Sanchez Medina}: Conceptualization, Data curation, Formal analysis, Investigation, Methodology, Project administration, Software, Validation, Visualization, Writing – original draft.
\textbf{Kai Sundmacher}: Funding acquisition, Supervision, Writing – review and editing.

\section*{Declaration of competing interest}
The authors declare that they have no conflict of interest.

\section*{Acknowledgments}
This work was partly supported by the Research In-itiative “SmartProSys: Intelligent Process Systems for the Sustainable Production of Chemicals”, funded by the Ministry for Science, Energy, Climate Protection and the Environment of the State of Saxony- Anhalt.

\section*{Declaration of generative AI and AI-assisted technologies in the writing process}
During the preparation of this work the authors used gpt-4o-2024-08-06 in order to check the grammar and improve the readability of the text. After using this tool, the authors reviewed and edited the content as needed and take full responsibility for the content of the published article.

\section{Appendix}
\appendix
\section{Derivation of the extended Margules model}
\label{app1}

The extended Margules model assumes that the molar excess Gibbs energy $g^E$ of any mixture can be described by a continuous function that is infinitely differentiable. Then, $g^E$ is expressed as a Taylor series, often sufficiently well-approximated when truncated after the third term. Therefore, for a mixture of $N$ components, we have that

\begin{align}
    g^E(\textbf{x}) = C_0 + \sum_{i=1}^{N-1} C_i x_i + \sum_{i=1}^{N-1} \sum_{j=1}^{N-1} C_{ij} x_ix_j + \sum_{i=1}^{N-1} \sum_{j=1}^{N-1} \sum_{k=1}^{N-1} C_{ijk} x_ix_jx_k
    \label{eq_taylor_ge}
\end{align}

\noindent where $\mathbf{x}=[x_1, x_2, \cdots, x_{N-1}]$ represents the vector of molar fractions for the $N-1$ components in the mixture, and $C$ denotes the polynomial coefficient corresponding to the indicated subscripts.

From the boundary condition given by Equation \ref{eq_boundary_condition}, it follows that $g^E=0$ when $x_N=1$. Therefore, $C_0=0$. Similarly, we know that $(C_i + C_{ii} + C_{iii})=0$ when $x_i=1 \quad \forall ~ i \in \{1,2, \dots, N-1\}$.

Therefore, Equation \ref{eq_taylor_ge} can be reformulated as:

\begin{align}
    g^E(\textbf{x}) = - \sum_{i=1}^{N-1} (C_{ii} + C_{iii}) x_i + \sum_{i=1}^{N-1} \sum_{j=1}^{N-1} C_{ij} x_ix_j + \sum_{i=1}^{N-1} \sum_{j=1}^{N-1} \sum_{k=1}^{N-1} C_{ijk} x_ix_jx_k
    \label{eq_taylor_ge2}
\end{align}

By leveraging the fact that the sum of all mole fractions adds to 1, we can re-express the first two terms of Equation \ref{eq_taylor_ge2} to obtain:

\begin{align}
    g^E(\textbf{x}) = - \sum_{i=1}^{N-1} \sum_{j=1}^{N} \sum_{k=1}^{N} (C_{ii} + C_{iii}) x_i x_j x_k + \sum_{i=1}^{N-1} \sum_{j=1}^{N-1} \sum_{k=1}^{N} C_{ij} x_ix_jx_k + \sum_{i = 1}^{N-1} \sum_{j = 1}^{N-1} \sum_{k = 1}^{N-1} C_{ijk} x_ix_jx_k
    \label{eq_taylor_ge3}
\end{align}

We can then simplify the expression by collecting all crossed terms to obtain a general expression that can be applied to estimate $g^E$ for a mixture of $N$ components:

\begin{align}
    g^E(\textbf{x}) = - \sum_{i=1}^{N-1} \sum_{j=i}^{N} \sum_{k=j}^{N} (\hat{C}_{ii} + \hat{C}_{iii}) x_i x_j x_k + \sum_{i=1}^{N-1} \sum_{j=i}^{N-1} \sum_{k=j}^{N} \hat{C}_{ij} x_ix_jx_k + \sum_{i = 1}^{N-1} \sum_{j = i}^{N-1} \sum_{k = j}^{N-1} \hat{C}_{ijk} x_ix_jx_k
    \label{eq_taylor_ge4}
\end{align}

For binary mixtures, Equation \ref{eq_taylor_ge4} simplifies to 

\begin{align}
    g^E = -\hat{C}_{111} x_1^2 x_2 + (-\hat{C}_{11} - \hat{C}_{111}) x_1 x_2^2
    \label{eq_ge_binary}
\end{align}

The value of the parameters $\hat{C}_{111}$ and $\hat{C}_{11}$ can be determined by considering the following relationship between $g^E$ and the IDACs:

\begin{equation}
    RT \ln{\gamma_{ij}^\infty} = \left( \frac{\partial g^E}{\partial x_i} \right)_{x_i \rightarrow 0, ~~ j \neq i} 
    \label{eq_determinar_constants}
\end{equation}

Therefore, the value of the constants is given by

\begin{align}
    \left( \frac{\partial g^E}{\partial x_1} \right)_{x_1 \rightarrow 0} &= - \hat{C}_{11} - \hat{C}_{111} = w_{12} = RT \ln{\gamma_{12}^\infty}\\
    \left( \frac{\partial g^E}{\partial x_2} \right)_{x_2 \rightarrow 0} &= - \hat{C}_{111} = w_{21} = RT \ln{\gamma_{21}^\infty}
\end{align}

\noindent and the final $g^E$ expression for the binary case results in Equation \ref{eq_ge_binary_final}.

The activity coefficients can be then calculated by differentiating Equation \ref{eq_ge_binary_final} in terms of the total excess Gibbs energy $G^E = M g^E$ (where, $M$ stands for the total number of moles) with respect to the number of moles of each component:

\begin{align}
    RT \ln{\gamma_1} =& \frac{\partial G^E}{\partial n_1} = w_{21} \left( 
    \frac{2 n_1 n_2}{M^2} - \frac{2 n_1^2 n_2}{M^3}
    \right)
    +
    w_{12}\left( 
    \frac{n_2^2}{M^2} - \frac{2 n_1 n_2^2}{M^3}
    \right) \\
    RT \ln{\gamma_1} =& 2 w_{21} x_1 x_2 + x_2^2 w_{12} - 2 g^E
\end{align}

\noindent and similarly for component 2:

\begin{equation}
    RT \ln{\gamma_2} = 2 w_{12} x_2 x_1 + x_1^2 w_{21} - 2 g^E
\end{equation}

A similar procedure can be followed to determined the expressions to compute the activity coefficients of components in a ternary mixture:

\begin{align}
    g^E = x_1 x_2 (x_2 w_{12} + x_1 w_{21}) + x_1 x_3 (x_3 w_{13} + x_1 w_{31}) + x_2 x_3 (x_3 w_{23} + x_2 w_{32}) + x_1 x_2 x_3 C_{123}
    \label{eq_ge_margules_ternary}
\end{align}

with $ C_{ijk} = \frac{1}{2}(w_{ij} + w_{ji} + w_{ik} + w_{ki} + w_{jk} + w_{kj}) - w_{ijk}$. While the binary parameters can be related to the corresponding IDACs as shown for the binary case, the ternary parameter $w_{ijk}$ is often set to zero as in \cite{andersen1981valid}. The resulting expressions for the activity coefficients are then given by:

\begin{equation}
    RT \ln{\gamma_1} = 2(x_1 x_2 w_{21} + x_1 x_3 w_{31}) + x_2^2 w_{12} + x_3^2 w_{13} + x_2 x_3 c_{123} - 2 g^E
\end{equation}

\begin{equation}
    RT \ln{\gamma_2} = 2(x_2 x_3 w_{32} + x_2 x_1 w_{12}) + x_3^2 w_{23} + x_1^2 w_{21} + x_3 x_1 c_{231} - 2 g^E
\end{equation}

\begin{equation}
    RT \ln{\gamma_3} = 2(x_3 x_1 w_{13} + x_3 x_2 w_{23}) + x_1^2 w_{31} + x_2^2 w_{32} + x_1 x_2 c_{312} - 2 g^E
\end{equation}




\bibliographystyle{elsarticle-num-names}
\bibliography{bibliography}

\end{document}